\documentclass{article}

\usepackage[utf8]{inputenc} 
\usepackage[T1]{fontenc}    
\usepackage{hyperref}       
\usepackage{url}            
\usepackage{booktabs}       
\usepackage{amsfonts}       
\usepackage{nicefrac}       
\usepackage{microtype}      
\usepackage{cleveref}       
\usepackage{lipsum}         
\usepackage{graphicx}
\usepackage{doi}
\usepackage{cite}
\usepackage{wrapfig}
\usepackage{lmodern}

\title{Mesons in nonlocal model with four-dimensional separable kernel}

\newif\ifuniqueAffiliation
\uniqueAffiliationtrue

\ifuniqueAffiliation 
\author{ A. ~Friesen$^1$, Yu. ~Kalinovsky$^1$, A. Khmelev$^{1}$ \\
	$^1$Joint Institute for Nuclear Research, Dubna, 141980\\ 
}

\begin{document}
\maketitle

\begin{abstract}
In this work we study the meson properties in the framework of an effective quark model. We start from the Bethe-Salpeter equation choosing the interaction kernel in nonlocal form with the Gaussian meson vertex function, characterized by a meson size parameter $\Lambda_H$. We demonstrate the model's predictive power by applying it to both light and heavy systems. Key results include a calculation of the $\pi^0\to\gamma\gamma$ decay width and the pion transition form factor $F_{\pi\gamma}(Q^2)$, which reproduces experimental data from low to high $Q^2$. We further predict the electromagnetic properties of heavy quarkonia, obtaining the two-photon decay widths of $\eta_c$ and $\eta_b$ and the radiative decay widths of $J/\psi$ and $\Upsilon$, all of which show consistency with available data and other theoretical approaches. The model provides a computationally efficient and unified framework for describing mesons from the light to heavy quark sectors.

\end{abstract} 



\section{Introduction}\label{intro}

A complete understanding of the full range of hadronic spectra from the lightest ($q\bar{q}$) to the heaviest ($Q\bar{Q}$) remains a challenge because it requires a unified treatment of chiral symmetry breaking and confinement at low energies together with heavy-quark dynamics at higher mass scales. For light flavors, spontaneous chiral symmetry breaking governs masses and couplings and underlies low-energy theorems such as GMOR, Goldberger–Treiman, and KSFR \cite{Schmidt:1995gea,Leupold:2003zb}. For heavy flavors, by contrast, chiral dynamics is subdominant. The strong dynamics of heavy quarks is simpler due to the emergence of an effective symmetry in the limit $m_Q \to \infty$, where they behave approximately as static color sources. The physics is well captured by heavy-quark spin symmetry, potential-model intuition \cite{Schmidt:1995gea,Leupold:2003zb}, lattice QCD \cite{ Burnier:2015tda} and nonrelativistic factorization ideas as formalized by nonrelativistic QCD (NRQCD) \cite{Bodwin:1994jh, Zalewski:1998gp}. A single framework that interpolates smoothly between these limits, while remaining predictive and computationally tractable, is therefore valuable both conceptually and phenomenologically.

 Effective quark models with nonlocal interactions provide such a bridge \cite{Branz:2009cd,Tran:2024phq,Dubnicka:2018gqg,Ivanov:1998ms} . Unlike local NJL-type constructions, nonlocal kernels yield ultraviolet-finite loop integrals and can be formulated to respect the vector and axial Ward – Takahashi identities \cite{Vogl:1991qt,Anikin:1996df}. This is essential for a consistent description of electromagnetic and axial currents and for reproducing anomaly-driven processes such as $\pi^0\rightarrow\gamma\gamma$. Models with nonlocal interaction are widely used to investigate light- and heavy- flavoured mesons \cite{Branz:2009cd,Tran:2024phq,Dubnicka:2018gqg,Ivanov:1998ms} at both zero and finite temperatures \cite{Blaschke:2000gd}.

In the present work we adopt a rank-1 separable Bethe - Salpeter kernel with a Gaussian vertex, which encodes the finite size of the meson through a single scale $\Lambda_H$ and ensures rapid convergence of loop integrals. The quark propagators are taken in Euclidean space with constituent masses that parametrize dynamical chiral symmetry breaking in the light sector and smoothly approach heavy-quark behavior for charm and bottom. The Bethe–Salpeter amplitudes are canonically normalized via the derivative of the polarization loop, guaranteeing consistent meson–quark couplings. Beyond vacuum properties, the same nonlocal structure can be extended to finite temperature and density \cite{Blaschke:2000gd}. Although our focus here is on $T = 0$, this portability highlights a practical advantage of the approach for connecting hadron structure to heavy-ion phenomenology in a controlled manner.

The Dyson–Schwinger/Bethe–Salpeter studies encode dynamical chiral symmetry breaking and have demonstrated unified descriptions of light and heavy mesons within a single formalism \cite{Ivanov:1998ms,PhysRevC.55.2649}.  Technically, we compute matrix elements with an alpha-representation for loop integrals, which reduces one- and three-point functions to low-dimensional integrals over scale variables and allows us to analyze a broad class of observables with minimal numerical overhead.

The parameters of the model consist of constituent quark masses and meson size scales $\Lambda_H$. The light masses and corresponding scale parameters $\Lambda_\pi, \Lambda_\rho$ are fitted by matching to a small set of well-measured observables, specifically light-meson masses and leptonic decay constants in the $\pi-\rho$ sector.  On the light side, we study the $\pi^0$ transition form factor in the space-like region and its two-photon decay, which test anomaly normalization and the approach to the perturbative QCD asymptote; for the decay width we compare to the PrimEx determination of $\Gamma(\pi^0\rightarrow\gamma\gamma)$ \cite{PrimEx:2010fvg}. On the heavy side, we fit the constituent masses of the heavier c- and b- quarks as well as $\Lambda_{\eta_c,\eta_b}$ constants on the base of masses and the leptonic decays of $\eta_c$ and $\eta_b$ mesons.  We compute the transition form factors and two-photon widths of $\eta_c$ and $\eta_b$ and confront them with available measurements and lattice benchmarks, including HPQCD’s determinations of heavy-quarkonia matrix elements \cite{Colquhoun:2023zbc} and the BaBar measurement of the space-like $\eta_c$ transition form factor \cite{BaBar:2010siw}. Radiative decays of vector mesons provide an additional, complementary probe: we analyze $\rho\rightarrow\pi\gamma$ as a light-sector test of current conservation and the momentum dependence of the vertex, and $J/\psi\rightarrow\eta_c\gamma$ and $\Upsilon\rightarrow\eta_b\gamma$  as heavy-quark spin–symmetry partners that test the model’s heavy-mass scaling.

The paper is organized as follows. In Section \ref{sec:Definition}, the basic tenets of the model are formulated. Section \ref{emagDec} is devoted to the transition form factors of the neutral pion and the heavy pseudoscalar mesons $\eta_c$ and $\eta_b$. Applications of the developed model to the radiative decays of the $\rho$ meson and heavy quarkonia are presented in Section \ref{Sec:VecMes}. The conclusions are provided in the final section. Technical details concerning the evaluation of one-loop integrals are given in Appendices A and B.

\section{Basic properties of pseudoscalar and vector mesons} \label{sec:Definition}
Mesons can be described as $q\bar{q}$ bound states using the Bethe-Salpeter equation, which in the ladder approximation has the form
\begin{equation}\label{eq:BS}
	\Gamma_H (q,P) = - \frac{4}{3} \int \frac{d^4 p}{(2\pi)^4}	D(q-p)\lbrace\gamma_\alpha S(p_1)\Gamma_H (p,P)	S(p_2) \gamma_\alpha\rbrace.  
\end{equation}
The vertex function $\Gamma_H (q,P)$ depends on the relative
four-momentum $q$, and the total four-momentum of the bound state $P$, $D(p-q)$ is the interaction kernel, $S(p)$ is the free propagator of the constituent quarks in Euclidean space.  In this paper we study the rank $-1$ separable  model with the interaction kernel \cite{Ivanov:1998ms,PhysRevC.55.2649}
\begin{equation}\label{eq:kernel}
	D(q-p) = D_0 \varphi(q^2) \varphi (p^2) ,
\end{equation}
where $D_0$ is the coupling constant and the function $\varphi
(p^2)$ is related to scalar part of the Bethe - Salpeter  vertex function. The vertex function  $\varphi(q^2)$  in the Gaussian form  $\varphi(q^2)= e^{-q^2/\Lambda_H^2}$ is used with the parameter $\Lambda_H$   characterizing the finite size of the meson. 

The basic object of our study is the meson vertex function, which with the separable Anzatz of the interaction kernel Eq.(\ref{eq:kernel}) has the form
\begin{equation}\label{eq:vert}
	\Gamma_H (p, P) = N_H \, \varphi (p^2) \,  \gamma_H \,.
\end{equation}
with  the normalization constant  $N_H$,  and the Dirac matrix $\gamma_H$ defined by the meson parity.  $S(p_i)$ is the dressed quark propagator in Euclidean space 
\begin{equation}\label{eq:q_prop}
	S(p_i)=\frac{1}{i (p_i\gamma) + m}.
\end{equation}
 
\begin{figure}
	\centerline{
		\resizebox{0.25\textwidth}{!}{%
			\includegraphics{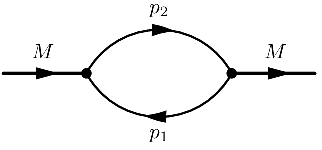}}
	}
	\caption{The meson polarization loop.}
	\label{fig:loop}       
\end{figure}

The meson-quark coupling constant $N_H$ is determined by the normalization condition for the Bethe-Salpeter amplitude. This condition involves the derivative of the polarization loop (presented in Fig. \ref{fig:loop}) with respect to the total momentum $P_\mu$, and is known as the compositeness condition \cite{Salam1962,PhysRev.130.776}:
\begin{equation}\label{eq:norm}
	1 = N_c \frac{ P_\mu}{2 P^2} \frac{\partial }{\partial P_\mu} 	\int \frac{d^4p}{(2\pi)^4} \mbox{tr}\left\lbrace 	\Gamma_H (p,P) S (p_1) \Gamma_H (q,P) S(p_2)   \right\rbrace
\end{equation}
evaluated at the on-shell point $P^2=-M_H^2$. Here $N_c=3$ is a number of colours, the momenta in quark propagators are defined as $p_i = p + q_i$, with $q_i = b_i P$, $b_1+b_2 = 1$.  A common choice for the momentum partitioning is $b_1 = -m_1/(m_1+m_2)$, $b_2 = m_2/(m_1+m_2)$, where $m_1$ and $m_2$ are the constituent masses of the quarks in the meson. 

Using the derivatives of the quark propagator Eq. (\ref{eq:q_prop}), we can write the normalization conditions Eq.(\ref{eq:norm}) for pseudoscalar mesons 
\begin{eqnarray}\label{Ps_norm}
	1 & = & -i \frac{N_c N_{ps}^2}{2 P^2}P^\mu \int\frac{d^4p}{(2 \pi)^4}\varphi^2 (p^2) \nonumber\\
	& \times& \lbrace 	b_1\mbox{tr} \left[	i\gamma_5 S(p_1)\gamma_\mu S(p_1) i\gamma_5 S(p_2) \right] + b_2\mbox{tr} \left[	S (p_1)i\gamma_5  S(p_2) \gamma_\mu  S(p_2) i\gamma_5\right] \rbrace,  
\end{eqnarray}
and for vector mesons:
\begin{eqnarray}\label{vec_norm}
	1 &= &-i \frac{N_c N_{\rm{v}}^2}{6 P^2}P^\mu \int\frac{d^4p}{(2 \pi)^4}\varphi^2 (p^2)\epsilon^{\rho\sigma}  \nonumber \\
	&\times& \lbrace b_1\mbox{tr} \left[	\gamma_\rho S(p_1)\gamma_\mu S(p_1) \gamma_\sigma S(p_2) \right]  + b_2\mbox{tr} \left[	S (p_1)\gamma_\rho S(p_2) \gamma_\mu  S(p_2) i\gamma_\sigma\right] 	\rbrace,  
\end{eqnarray}  
where the factor 1/3 appears because the three transverse directions are summed.  Here $\epsilon^{\mu\nu} = g^{\mu\nu}-P^\mu P^\nu/P^2$ is the polarization tensor for the vector meson satisfying  $\epsilon^{\mu\nu}P^\nu = 0$ and $\epsilon^{\mu\nu}\epsilon_{\mu\nu}= 3$.

For pseudoscalar mesons the weak leptonic decay constant can be obtained from the matrix element of the axial current 
\begin{equation}
	f_{p} p^\mu = \langle 0| \bar{q_a} \gamma_\mu\gamma_5 {q_a}|P(p)\rangle ,
\end{equation}
where $a, b$ are the quark flavours, $a,b = (u, d, s, c)$ and $q_a, q_b$ correspond to the quark content of the meson, in integral form this relation has view:
\begin{equation}\label{Ps_const} 
	P^\mu f_{p} = N_c N_{p}\int\frac{d^4p}{(2 \pi)^4}\varphi (p^2) \rm{tr} \lbrace (i \gamma_5) S (p_1)(\gamma_\mu \gamma_5) S(p_2)\rbrace. 
\end{equation}
The vector leptonic decay can be written in the same way \cite{Ivanov:1998ms}:
\begin{equation}
	f_{\rm v} M_{\rm v} \epsilon^{\mu\nu}= \langle 0| \bar{q_a}\gamma^\mu q_b |V_\nu(p)\rangle
\end{equation}
which in integral form can be expressed as a loop integral \cite{Gasser:1987,Maris:PhysRevC60}
\begin{equation} 
	f_{\rm{v}}  M_{\rm_v}\epsilon^{\mu\nu} = N_c N_{\rm{v}}\int\frac{d^4p}{(2 \pi)^4}\varphi (p^2) \rm{tr} \lbrace \gamma_\mu  S(p_1)\gamma_\nu S(p_2) 	\rbrace. 
\end{equation}
To determine the decay constant for a multi-flavour configuration state such as $\rho^0$, it is better to use the electromagnetic decay coupling, which can be conventionally expressed via the dimensionless coupling constant $g_{\rm v}$ in the form  $ f_{\rm{v}}= g_{\rm{v}} M_{\rm{v}}$, where for the $\rho$ meson the factor $\sqrt{2}$ also appears due to the meson flavour structure \cite{Gasser:1987,Maris:PhysRevC60}.

The system of equations that defines the model parameters is complete. The free parameters of this model are the constituent quark masses $m_u=m_d, m_s, m_c$ and the set of the state-specific scale parameters $\Lambda_{\rm{H}}$. Their values are determined by fitting experimental observables, such as the meson masses and decay constants. The normalization conditions in Eqs.
(\ref{Ps_norm}) - (\ref{vec_norm}) yield the constants $N_{\rm{v}}$, $ N_{p}$,  which represent the meson-quark coupling constants for vector and pseudoscalar states, respectively. The one-loop integrals resulting from taking the trace in the above equations  are detailed in Appendix A. Using the experimental values $m_\pi = $ 0.139 GeV, $f_\pi =  $0.131 GeV,  $m_\rho = $ 
0.77 GeV, $f_\rho =  $0.2 GeV \cite{Maris:PhysRevC60,Panda:bgjp23} along with the explicit Gaussian form of the BS amplitude $\varphi(p^2) = e^{-p^2/\Lambda_H^2}$, we fitted the light quark mass and parameters $\Lambda_\pi, \Lambda_\rho$ and calculated the coupling constants $N_\pi, N_\rho$ for light mesons. The resulting parameters are summarized in Table \ref{tab:params1}.
\begin{table}[h]
\centering
	\caption{Basic model parameters and constants for light mesons.}
	\label{tab:params1}       
	\begin{tabular}{ccccc}
		\hline\noalign{\smallskip}
		$m_{u(d)}$, GeV& $\Lambda_\pi$, GeV&$N_\pi$&$\Lambda_\rho$, GeV&$N_\rho$  \\
		\noalign{\smallskip}\hline\noalign{\smallskip}
		0.223& 1.14&3.724&0.75&3.612\\
		\noalign{\smallskip}\hline
	\end{tabular}
\end{table}

\begin{figure}
	\centerline{
		\resizebox{0.25\textwidth}{!}{%
			\includegraphics{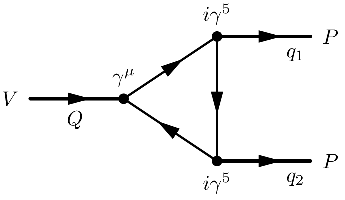}}
	}
	\caption{The diagrams of $\rho\rightarrow\pi\pi$ decay.}
	\label{fig:VPP}       
\end{figure}

To test the model, we study the hadronic decay of the  $\rho$-meson into two pions  $\rho\rightarrow\pi\pi$. The Feynman diagram for the $\rho\pi\pi$   decay is presented in Fig.~\ref{fig:VPP} and the corresponding matrix element is given by
\begin{equation}\label{eqn:vpp}
	T^\mu(q_1,q_2) = N_c N_\rho N_\pi\int \frac{d^4p}{(2\pi)^4}	\varphi(p^2) \mbox{tr}\lbrace  \gamma_\mu S(p_2) i \gamma_5 	S(p_3) i \gamma_5  S(p_1) \rbrace,
\end{equation}
where  $\varphi(p^2)= \varphi_\rho(p^2)\varphi_\pi^2(p^2)$. The matrix element can be decomposed into two terms:  
\begin{equation}
	T^\mu(p_1, p_2) = (p_1-p_2)^\mu f^+(t) + (p_1+p_2)^\mu f^-(t),
\end{equation}
with  $t = -(p_1-p_2)^2$. The physical constraints are $f^-(t = M_\rho^2) = 0$, $\frac{1}{2}f^+(t = M_\rho^2) = g_{\rho\pi\pi}$. 

The calculation technique for the one-loop  three-point  integrals is detailed in Appendix B. The  decay width for the $\rho\rightarrow\pi\pi$ decay is calculated as
\begin{equation}
	\Gamma_{\rho\pi\pi} = \frac{1}{6 \pi} \frac{k^3}{M_\rho^2} g_{\rho\pi\pi}^2,
\end{equation}
where $\displaystyle{k =2\left(M_\rho^2- 4 M_\pi^2\right)^{1/2}}$ is the pion momentum in the $\rho$-meson rest frame and the factor  1/3 appears because the three transverse directions are summed. Using the parameters from Table \ref{tab:params1}, we obtain  $g_{\rho\pi\pi} = 6.08$ and $\Gamma_{\rho\pi\pi} =0.151$ GeV. This results is in good agreement with the Lattice QCD calculations $g_{\rho\pi\pi} =$ 5.69(13) and $\Gamma_{\rho\pi\pi} = 0.155$ GeV \cite{Alexandrou:2017mpi}, and with the experimental value for $\rho^+\rightarrow\pi^0\pi^+$ of $\Gamma_{\rho\pi\pi} = 0.150\pm 0.005$ GeV \cite{PhysRevD.33.3199}.

\section{Electromagnetic decay of light and heavy pseudoscalar mesons} \label{emagDec}

This Section is devoted to discussion of the electromagnetic decay and transition form factors for neutral pion and heavy $\eta_c$ and $\eta_b$ mesons. The study of electromagnetic decays is important for understanding of the structure and binding mechanisms of hadrons and provide a good test for applicability of the suggested approach.

\begin{figure}
	\centerline{
		\resizebox{0.25\textwidth}{!}{%
			\includegraphics{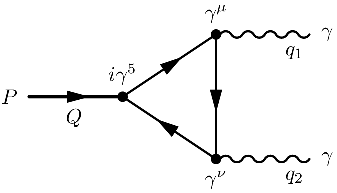}}
	}
	\caption{The diagram of electromagnetic $P\rightarrow\gamma\gamma$ decay.}
	\label{fig:Pgg}       
\end{figure}

The decay amplitude of the process $P \rightarrow\gamma \gamma$ is described by triangle Feynman diagram in Fig. \ref{fig:Pgg} and reads as
\begin{equation}
	T^{\mu\nu}(q_1, q_2) = 	N_c N_p \int \frac{d^4p}{(2\pi)^4}\varphi(p^2)\mbox{tr}\lbrace i\gamma_5 S(p_2)   	\gamma_\mu S(p_3) \gamma_\nu S(p_1) \rbrace 
\end{equation}

After applying the pion vertex function and taking the trace, the amplitude can be written as
\begin{eqnarray}\label{eq:pvv}
	T(q_1, q_2) &=&
	i \epsilon_{\mu \nu \alpha \beta } \epsilon_1^\mu \epsilon_2^\nu
	q_1^\alpha q_2^\beta \left( N_c Q_q^2 \right)
	\frac{m N_\pi }{ 4 \pi^2} I(Q,q_1,q_2) \nonumber \\
	&=& i \epsilon_{\mu \nu \alpha \beta } \epsilon_1^\mu \epsilon_2^\nu
	q_1^\alpha q_2^\beta \left( N_c Q_q^2 \right)
	G_{\pi\gamma\gamma}(Q, q_1, q_2) ,
\end{eqnarray}
where  $q_1, q_2$ are momenta of photons, $Q_q$ is quark charge, for case with pion $Q_q^2 = (e_u^2-e_d^2)$ and $e_u=2/3e$, $e_d=-1/3e$. 

\begin{figure*}
\centerline{
\includegraphics[width=0.47\textwidth]{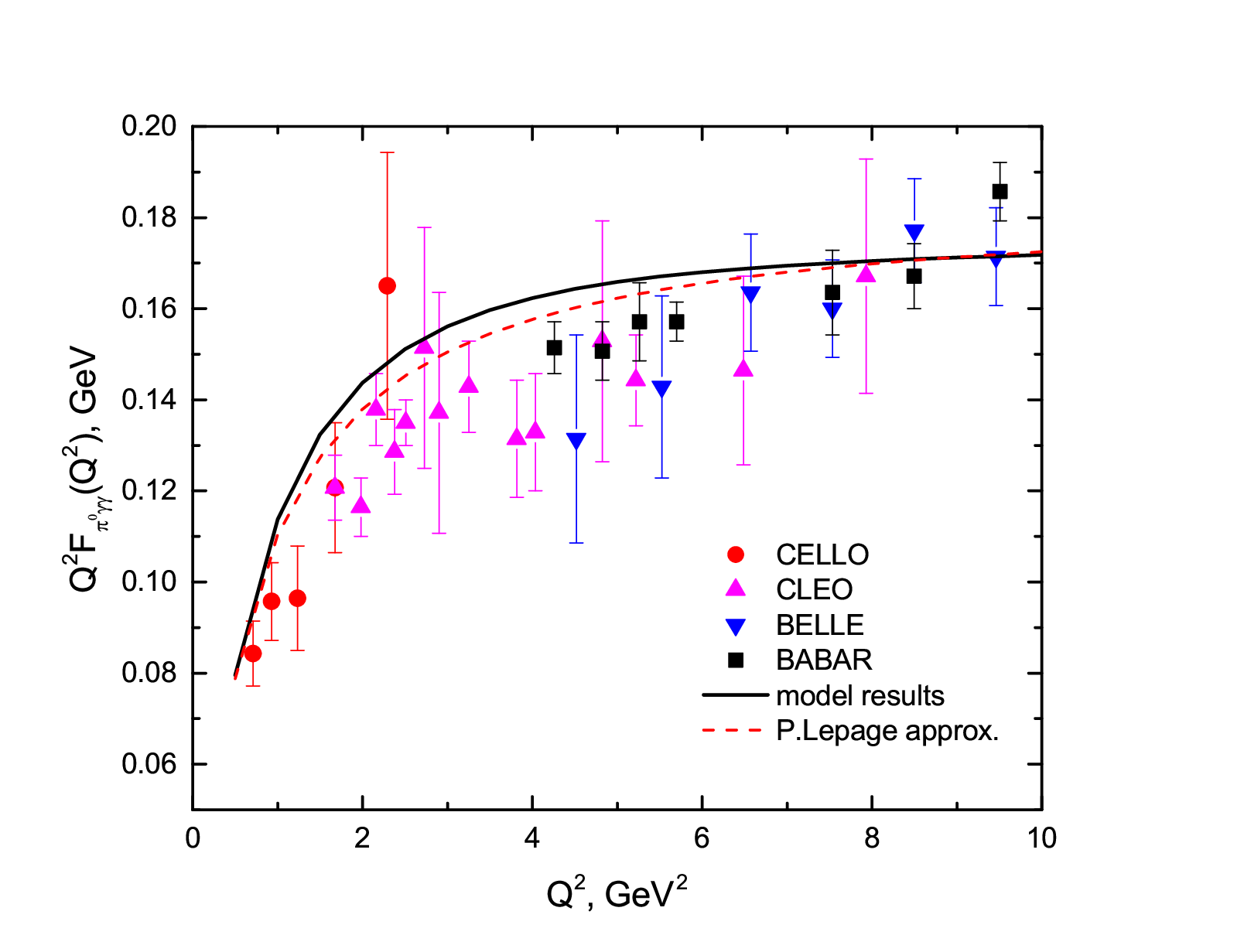}
\includegraphics[width=0.47\textwidth]{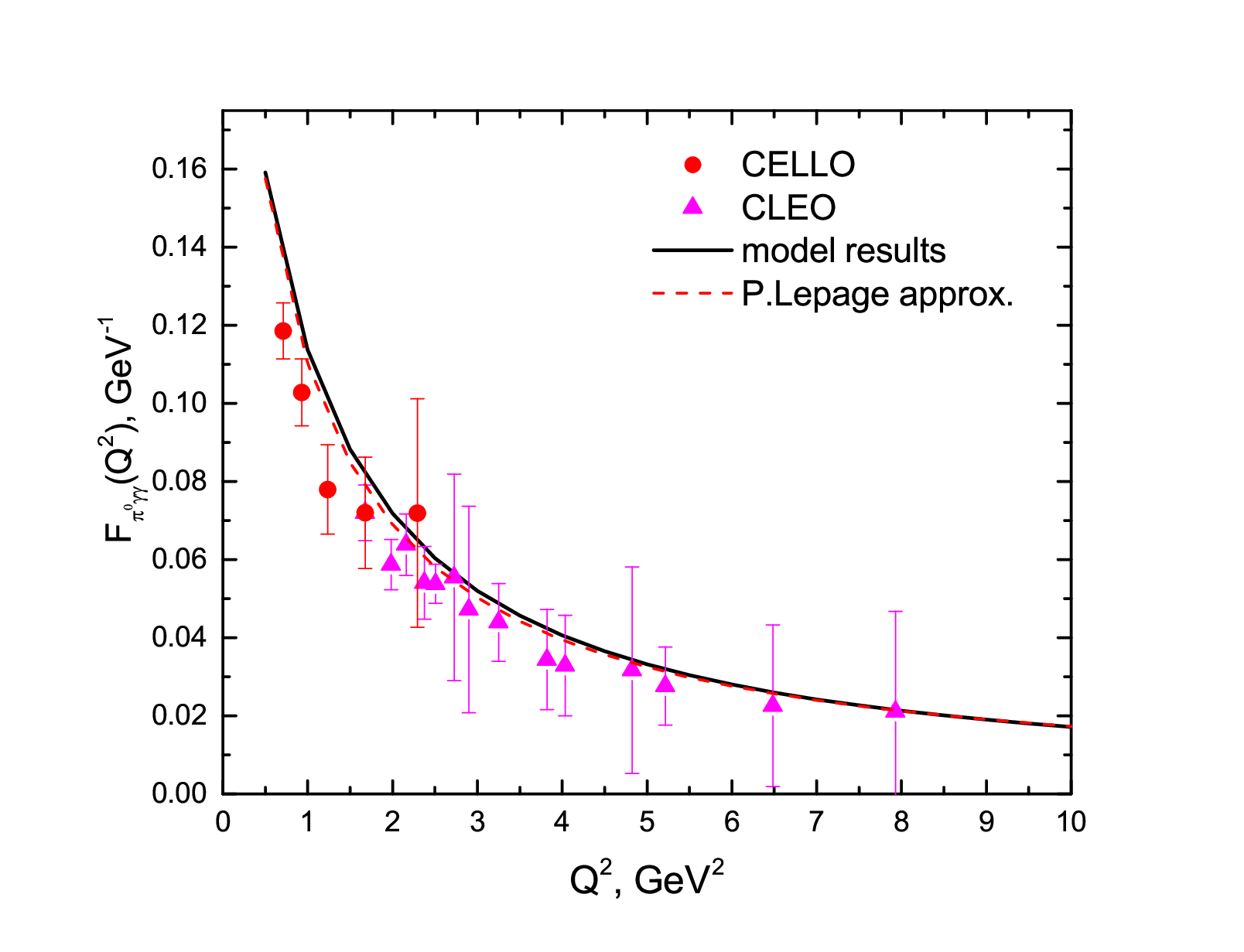}}
	\caption{The transition form factor $F_{\pi\gamma}$ as function of the space-like photon momentum $Q^2$. Experimental data are taken from \cite{BaBar:2009rrj,Belle:2012wwz,CLEO:1997fho,CELLO:1991zPh}.}
	\label{fig:piTFF}       
\end{figure*}

The two - photon decay coupling constant in our model is
obtained from Eq. (\ref{eq:pvv}) for two on-shell photons, when $q_1^2$ and $q_2^2$ are equal to zero.
\begin{equation}
	g_{\pi \gamma \gamma} = G_{\pi\gamma\gamma}(M_\pi^2,0,0) \simeq
	\frac{m N_\pi }{ 4 \pi^2 \Lambda_\pi^2} I(M_\pi^2, 0, 0),
\end{equation}
then the decay width can be considered as
\begin{equation}\label{eq:Gpigg}
	\Gamma (\pi^0 \rightarrow \gamma\gamma) = \frac{M_\pi^3}{64 \pi} (4\pi\alpha)^2 g_{\pi\gamma\gamma}^2.
\end{equation}

The transition form factor with one of photons being off-shell ($q_2^2 = -Q^2$) is defined as
\begin{equation}\label{eq:piggFF}
	F_{\pi \gamma} (Q^2) = e^2 G_{\pi \gamma\gamma}(M_\pi^2,Q^2,0).
\end{equation}
For the model parameters in Table.~\ref{tab:params1}  the width $\Gamma_{\pi\gamma\gamma} = 7.76$ eV was obtained, and from the last experiment it is known that $\Gamma^{\rm exp}_{\pi\gamma\gamma} = 7.82$ eV \cite{PrimEx:2010fvg}.

The pion electromagnetic decay has been the subject of extensive experimental and theoretical study. Experimentally, the form factor $F_{\pi\gamma}(Q^2)$ has been measured at $e^+ e^-$ colliders by the CELLO, CLEO, BaBar and Belle  Collaborations\cite{CELLO:1991zPh,CLEO:1997fho, BaBar:2009rrj, Belle:2012wwz}. The latter two extended the momentum transfer range from 4 to 40 GeV$^2$ and observed an unexpected growth with increasing $Q^2$. The first theoretical estimation of the pion transition form factor as function of the transferred momentum was done by S. Brodsky and P. Lepage in the framework of non-perturbative QCD  \cite{PhysRevD.24.1808}:
\begin{equation}
	F_{\pi\gamma}(Q^2) = \frac{2 f_\pi}{8\pi^2f_\pi^2 + Q^2},
	\label{TFF_prediction}
\end{equation}
which predicted limit $Q^2F_{\pi\gamma} = 2 f_\pi$ at $Q^2\rightarrow\infty$.
The transition form factor $F_{\pi \gamma}(Q^2)$ as function of $Q^2$  for our model is shown in Fig.~\ref{fig:piTFF}.  The behavior of form factor demonstrates good agreement to the experimental data and the classic limit case Eq.~(\ref{TFF_prediction})  (red dashed line).

For calculation of the two-photon decay widths of the $\eta_c$ and $\eta_b$ the quark masses $m_c = 1.6$ GeV and $m_b = $ 4.77 GeV and the scale parameters $\Lambda_H$ were fitted to the masses of mesons and decay constants. The values used to fit and resulting constants $N_H$ are presented in Table~\ref{tab:heavyps}. 

\begin{table}[h]
\centering
	\caption{Model parameters and constants for heavy pseudoscalar mesons $\eta_c$ and $\eta_b$.}
	\label{tab:heavyps}       
	\begin{tabular}{cccccc}
		\hline\noalign{\smallskip}
		&$M_H$, GeV&$\Lambda_H$, GeV&$N_H$&$f_H$, GeV& $f_H$, GeV\\
			&&&& (our)&  (refs)\\
		\noalign{\smallskip}\hline\noalign{\smallskip}
		$\eta_c$&2.985&2.775&3.546&0.426& 0.42\cite{Hwang:1997ie} \\
		$\eta_b$ &9.39&2.81&8.568&0.715& 0.705\cite{Hwang:1997ie}\\
		\noalign{\smallskip}\hline
	\end{tabular}
\end{table}

The decay widths $\Gamma_{\eta_{c(b)}\gamma\gamma}$ and the corresponding transition form factors were calculated using  Eq. (\ref{eq:Gpigg}) and Eq.(\ref{eq:piggFF}). The quark charge $Q_q^2$ was set to $(4/9)e^2$ for the charm quark and $(1/9)e^2$ for the bottom quark. With the chosen parameter set, the calculated width for $\eta_c\to\gamma\gamma$ is $\Gamma_{\eta_c\gamma\gamma} = 5.03$ keV. This result compares with available Lattice QCD data, which span  from 4.88 keV \cite{Ryu:2018egt} to 6.788 keV \cite{Colquhoun:2023zbc}, while the PDG  average shows 5.1$\pm 0.4$ keV. For the $\eta_b$ meson we obtained the result $\Gamma_{\eta_b\gamma\gamma} = 0.18$ keV, which lies on the lower end of predictions from other models, which range from 0.17 keV in \cite{Ackleh1992} to 0.659$\pm$92 keV \cite{Penin:2004ay} (for a comprehensive review, see ref.\cite{Lansberg:2006sy}).

The theoretical and experimental situation regarding the study of the transition form factors and electromagnetic decays for $\eta_c$ and $\eta_b$ is less established.  For the charmonium case, the form factor $F_{\eta_c \gamma} (Q^2)$ was measured by the BaBar collaboration in the space-like region up to 50 GeV$^2$ \cite{BaBar:2010siw}. Theoretically $\eta_c$ TFF has een studied in the space-like region within frameworks pQCD \cite{Feldmann:1997te}, the light-front quark model \cite{Ryu:2018egt}, and a covariant approach based on Dyson-Schwinger and Bethe-Salpeter (BS)\cite{Chen:2016bpj}.

\begin{figure}[h]
	\centerline{
		\includegraphics[width=0.5\textwidth]{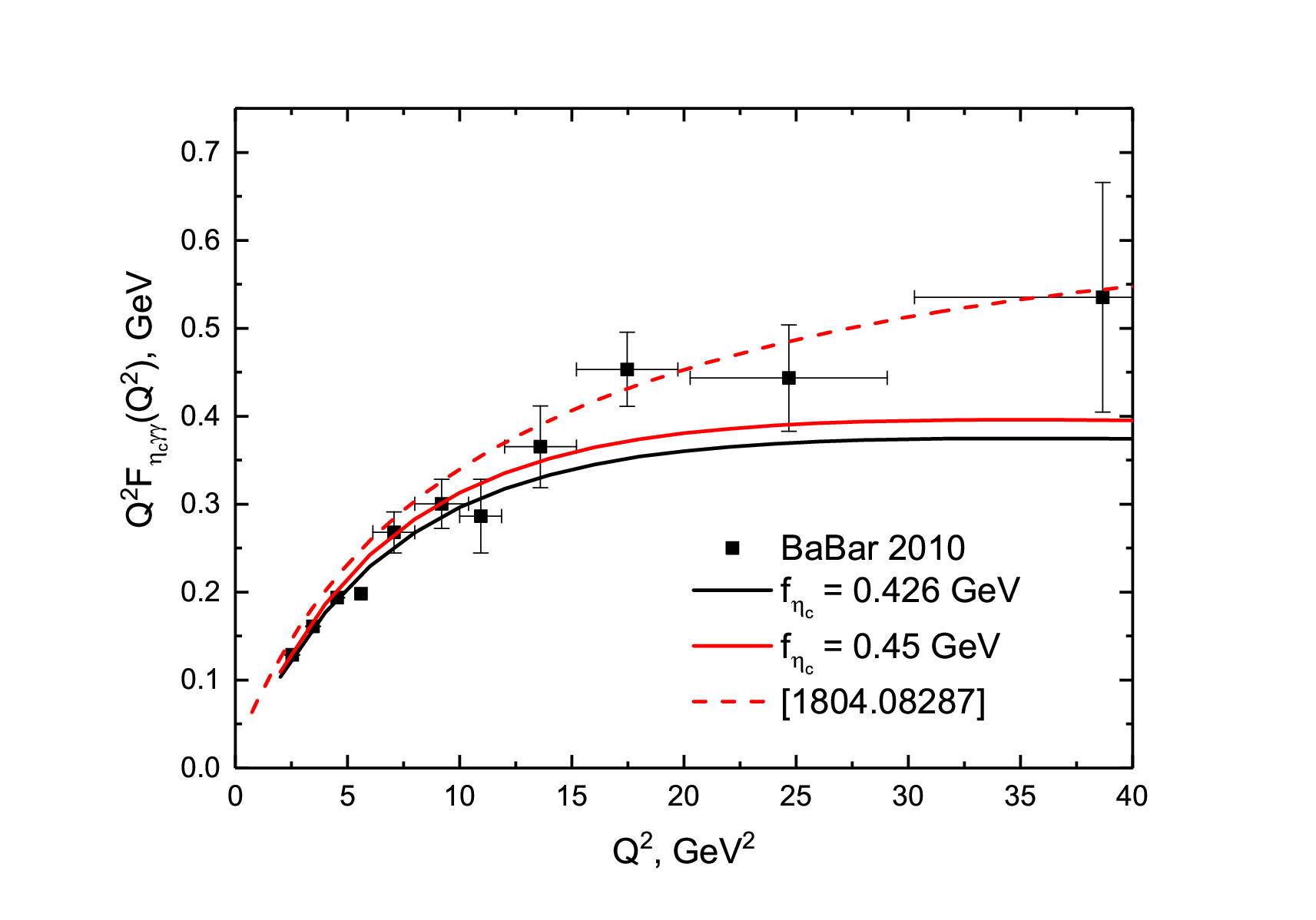}
		\includegraphics[width=0.47\textwidth]{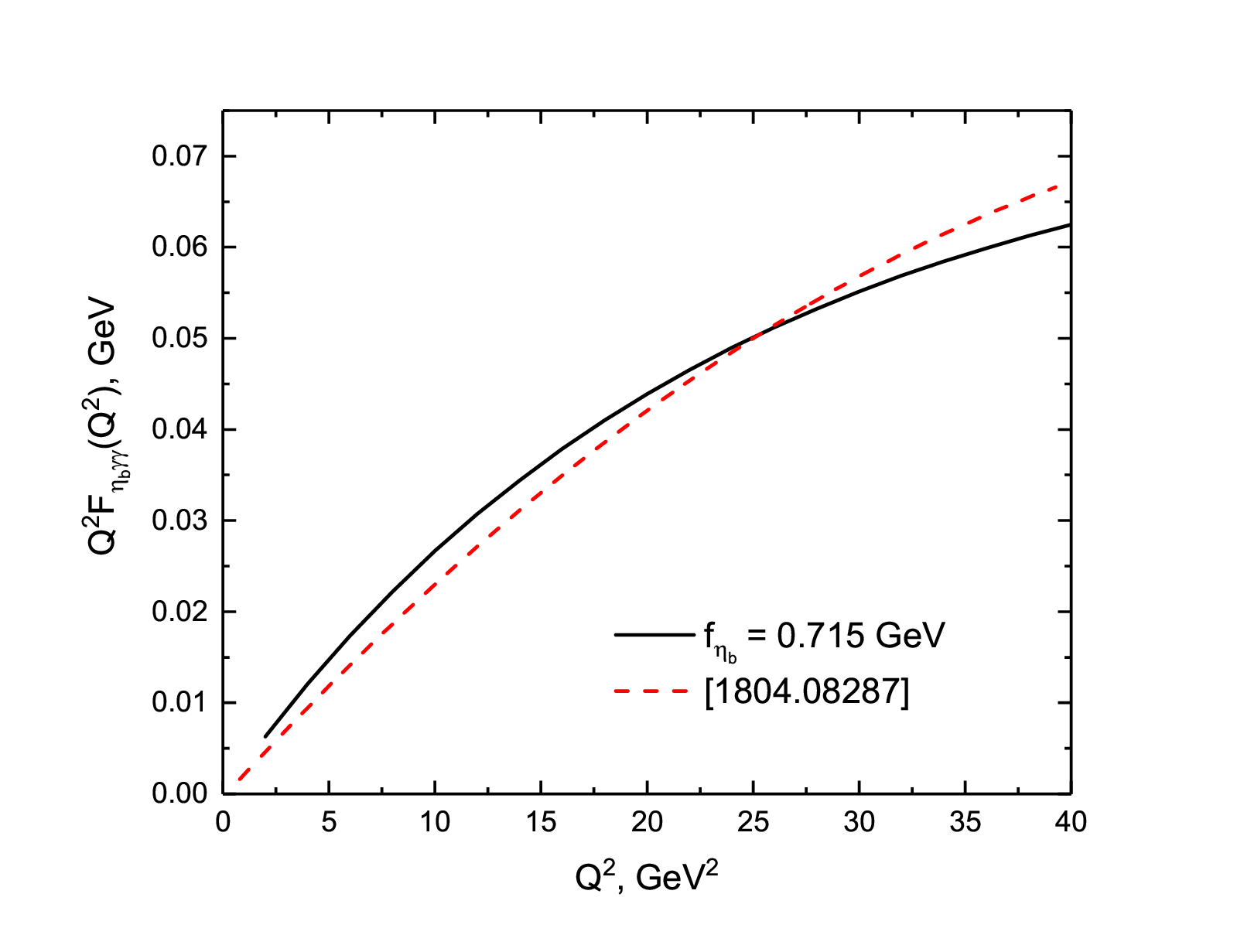}
	}
	\caption{Form factor $F_{\eta_c\gamma}$ and $F_{\eta_b\gamma}$ as function of the space-like photon momentum $Q^2$. Data are taken from \cite{BaBar:2010siw} for $\eta_c$ and from the light-front quark model for $\eta_b$\cite{Ryu:2018egt}.}
	\label{fig:formfactor_cb}       
\end{figure}

The left panel of Fig.~\ref{fig:formfactor_cb} shows the transition form factor $F_{\eta_c\gamma}$ for two values of decay constant $f_{\eta_c} = 0.426$ GeV and $f_{\eta_c} = 0.450$ GeV compared to the BaBar experimental data \cite{BaBar:2010siw}.  Due to the lack of experimental data and limited theoretical studies for the $\eta_b$, the right panel of Fig.~\ref{fig:formfactor_cb} compares our predictions for the transition form factor $F_{\eta_b\gamma}$ with the results of the light-front quark model \cite{Ryu:2018egt}. As shown, the best agreement between our model of the experiments appears at momenta $Q^2<15$ GeV$^2$.

\section{Radiative decays of vector mesons} \label{Sec:VecMes}

\begin{figure}[t]
	\centerline{
		\resizebox{0.25\textwidth}{!}{%
			\includegraphics{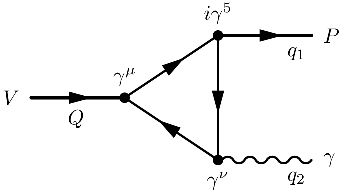}}
	}
	\caption{The  Feynman diagram of $V\rightarrow P\gamma$ radiative decay.}
	\label{fig:VPg}       
\end{figure}

In this Section radiative decays of vector mesons into pseudoscalar mesons and photons are discussed. The  decay ($V\rightarrow PV$) is defined by the Feynman diagram in Fig. \ref{fig:VPg} and the  corresponding matrix element is given by
\begin{equation}\label{eqn:vpv}
	T^{\mu\nu}(q_1,q_2) = 
	N_c N_{\rm v}N_p\int \frac{d^4p}{(2\pi)^4}\varphi(p^2)
	\mbox{tr}\lbrace \gamma_{\nu} S(p_2)
	i\gamma_5 S(p_3) \gamma_\mu S(p_1) \rbrace,
\end{equation}
where $\varphi(p^2) = \varphi_{\rm v}(p^2)\varphi_p(p^2)$. Then
\begin{equation}\label{eq:ropig}
	T^{\mu\nu}(q_1, q_2) =
	i \epsilon_{\mu \nu \alpha \beta } \epsilon^\mu(Q) \epsilon_2^{*\nu}(q_2)
	Q^\alpha q_2^\beta \left( N_c Q_qe \right)\frac{m N_p N_{\rm v} }{ 4 \pi^2} I(Q,q_1,q_2). 
\end{equation}

The decay width is defined as
\begin{equation}\label{eqn:Gvpv}
	\Gamma_{VP\gamma} = \frac{1}{3}\alpha k^3 g_{VP\gamma}^2,
\end{equation}
where 1/3 appears due to  vector polarizations, and factor $k = (M_{\rm v}^2-M_p^2)/2M_{\rm v}$.

The formalism is applied to calculate the decay width for the light $\rho$ meson and then to the heavy quarkonia states $J/\psi$ and $\Upsilon$. Using model parameters for the light quark and $\rho$ meson from Table \ref{tab:params1}  the radiative decay width for the light $\rho$ meson was calculated to be  $\Gamma_{\rho\pi\gamma} =71.8$ keV. This result is in consistent with the Lattice QCD result $\Gamma_{\rho\pi\gamma} = 84.2(6.7)(4.3)$ keV\cite{Leskovec:2018wuo} and experimental value of 59.814 keV  \cite{PhysRevD.33.3199}. The properties of heavy vector mesons are calculated with the same quark masses as in previous Section, namely $m_c = 1.6$ GeV and $m_b = $ 4.77 GeV. The masses of the mesons and decay constants used for the fit as well as size parameters $\Lambda_H$ and coupling constants $N_H$ are presented in Table~\ref{tab:heavymesons}. 

\begin{table}[h]
\centering
	\caption{Model parameters and constants for heavy quarkonia.}
	\label{tab:heavymesons}       
	\begin{tabular}{cccccc}
		\hline\noalign{\smallskip}
		&$M_H$, GeV&$\Lambda_H$, GeV&$N_H$&$f_H$, GeV& $f_H$, GeV \\
		&&&&(our)& (refs)\\
		\noalign{\smallskip}\hline\noalign{\smallskip}
		$J/\psi$&3.075&2.03&3.238& 0.435&0.399 \cite{Dudek:2006ej}\\
		$\Upsilon$&9.405&4.22&3.489& 0.705& - \\
		\noalign{\smallskip}\hline
	\end{tabular}
\end{table}

\begin{table}[h]
\centering
	\caption{The radiative decay widths for heavy mesons.}
	\label{tab:f_vpv}       
	\begin{tabular}{ccc}
		\noalign{\smallskip}\hline\noalign{\smallskip}
		& our & refs. \\
		\noalign{\smallskip}\hline\noalign{\smallskip}
		$J/\psi\rightarrow\eta_c\gamma$& 2.28 keV&1.83-2.49 keV (see text)\\
		$\Upsilon\rightarrow\eta_b\gamma$& 25.3 eV& 5.8 \cite{Ebert:2002pp}, 45 eV \cite{Choi:2007us}\\
		\noalign{\smallskip}\hline
	\end{tabular}
\end{table}

To calculate the decay widths of the radiative decay processes $J/\psi\rightarrow\eta_c\gamma$, $\Upsilon\rightarrow\eta_b\gamma$, the appropriate vertices, masses, and quark charges must be substituted into Eq.~(\ref{eqn:Gvpv}). Theoretical predictions for the $\Upsilon\rightarrow\eta_b\gamma$ vary significantly for different models. The light-front quark model yields $\Gamma_{\Upsilon\eta_b\gamma} = 45$ eV  \cite{Choi:2007us}, while the relativistic quark model predicts $\Gamma_{\Upsilon\eta_b\gamma} = 5.8$ eV  \cite{Ebert:2002pp}. For the  $J/\psi\rightarrow\eta_c\gamma$  decay width the Lattice QCD provides robust theoretical results, with calculations yielding  $\Gamma_{J/\psi\eta_c\gamma} = 2.219$ keV  \cite{Colquhoun:2023zbc} and $\Gamma_{J/\psi\eta_c\gamma} = 2.49$ keV \cite{Donald:2012ga}. These can be compared to the experimental data in the CLEO experiment $\Gamma_{J/\psi\eta_c\gamma}^{\rm exp} = 1.83$ keV \cite{CLEO:2008pln}  and in the KEDR experiment $\Gamma_{J/\psi\eta_c\gamma} = 2.17$ keV \cite{Anashin:2010nr}. The results obtained within our model for these decays are summarized in Table \ref{tab:f_vpv}.

\section{Conclusions}

In this work, we develop an approach for describing the properties of both light and heavy mesons within the framework of an effective quark theory with nonlocal interaction. Such nonlocal models are widely used to study light- and heavy-flavored mesons \cite{Branz:2009cd,Tran:2024phq,Dubnicka:2018gqg,Ivanov:1998ms} at both  zero and finite temperature \cite{Blaschke:2000gd}. In this model the nonlocality of quark-antiquark interaction is implemented through the four - momentum Gaussian form factor in the meson vertex function. The model parameters include the constituent quark masses and the meson size parameters $\Lambda_H$ and the normalization constants $N_H$, which play a role of meson-quark coupling, are obtained in explicit form. All parameters are determined by fitting to physical observables such as electromagnetic and leptonic decay constants. 

To test of the model's applicability, we calculated physical processes: the two-photon decay of pseudoscalar mesons and the radiative decay of vector mesons in pseudoscalar and photon.  The results show good agreement with experimental data and with predictions from other models.

A key advantage of this model is the simplicity of evaluating one-loop integrals. Further technical details required for analytical calculations are provided in the Appendices. Looking forward, this work provides a solid foundation for several exciting extensions. These include the study of mesons with open flavor (e.g., $D, B$), their excited states, and the application of the model to explore the properties of hadrons in extreme environments of finite temperature and density, relevant to heavy-ion collision experiments.


\section*{Appendix A: integration techniques}

In order to demonstrate how the loop integrations in this work are carried out, let's start from the one-loop two-point integral: 
\begin{equation}\label{twopoint}
I (P^2) = \int \frac{d^4p}{\pi^2} F(p^2)\frac{1}{ \left( p_1^2+m_1^2\right)
	\left( p_2^2+m_2^2	\right)	}, 
\end{equation}
where $p_1= p +q_1$ and $p_2= p + q_2$ with 
$q_i = b_i P$ and $b_1 = - m_1/(m_1+m_2)$, $b_2 = m_2/(m_1+m_2)$, $b_2+b_1 = 1$. The function $F(p^2)$ contains the vertex functions $\varphi(p^2/\Lambda_H^2)$. 
With the help of the Feynman parametrization we can write (\ref{twopoint}) as 
\begin{equation} \label{eq_after_FP}
I (P^2) = \int \frac{d^4p}{\pi^2} F(p^2) \int_0^1 d\alpha_1\frac{1}{(\alpha_1(p_1^2+m_1^2)+\alpha_2(p_2^2+m_2^2))^2}
\end{equation}
with $\alpha_2 = 1-\alpha_1$.

The denominator in Eq. (\ref{eq_after_FP}) is presented as 
\begin{eqnarray}
\alpha_1 (p^2+2(pq_1)+q_1^2+m_1^2)+
	\alpha_2 (p^2+2(pq_2)+q_2^2+m_2^2)  = (p+R)^2+D	
\end{eqnarray}
where $\displaystyle R = \sum_i \alpha_i q_i$ and 
$\displaystyle D = \sum_i{\alpha_i(q_i^2+m_i^2)} - R^2$.
It gives 
\begin{eqnarray}\label{eq:simpl}
	I (P^2) = \int [d \alpha] \int \frac{d p}{\pi^2} F(p^2)  \frac{1}{((p+R)^2+D)^2}\, .
\end{eqnarray}
where 
$\displaystyle \int [d \alpha] = \int_0^1 d \alpha_1 
\int_{0}^{1-\alpha_1} d\alpha_2
\delta(\alpha_1+\alpha_1-1)$.

Using  the integral representation in Euclidean space
\begin{eqnarray*}
	\frac{1}{((p+R)^2+D)^n} =\frac{1}{\Gamma(n)} \int t^{n-1} dt \,\, e^{-t ( (p+R)^2+D )}\,.
\end{eqnarray*}
and the Laplace transformation for the function $F(p^2)$: 
\begin{eqnarray}
	\int s^{n} ds \tilde{F}(s) e^{-s z_0} = (-1)^nF^{(n)}(z_0)
\end{eqnarray}
the main integral Eq.(\ref{twopoint}) can be written as 
\begin{eqnarray}
I (P^2) = 	\int [d\alpha ] ds \tilde{F} (s) t dt \int \frac{d p}{\pi^2}
	e^{-s p^2} e^{-t ( (p+R)^2+D )}
\end{eqnarray}
Taking into account the relation
\begin{eqnarray}\label{add_repl}
	s p^2 + t ( (p+R)^2+D ) = (s+t) \left(p + \frac{t}{s+t}R \right)^2 +t D + \frac{st}{s+t} R^2 \, ,
\end{eqnarray}
and changing the variables $t = st$ 
and using the inverse Laplace transformation:
\begin{eqnarray} \label{result}
	I (P^2) &=& \int [d\alpha ] \int ds \tilde{F} (s)\int t dt 
	e^{-s ( t D + \frac{t}{1+t}R^2)}
	\int \frac{d p}{\pi^2} e^{-s(1+t)p^2}  
	\nonumber \\
	&=&\int_0^1 d\alpha \int_0^{\infty}dt \frac{t}{(1+t)^2}  \left[ F(z_0)\right]\, 
\end{eqnarray}
with 
\begin{eqnarray}\label{eq:z0}
	z_0 &=& t D + \frac{t}{1+t}R^2.
 \end{eqnarray}

In general, after taking the traces during the Feynman diagram evaluation, not only scalar integrals appear. The most common expression for such integrals can be written as
\begin{equation}
    \int\frac{d p}{\pi^2}F(p^2)\frac{ \lbrace p^\mu, p^\mu p^\nu, p^\mu p^\nu p^\rho, \ldots \rbrace}{\left[p_1^2+m_1^2\right]
	\left[ p_2^2+m_2^2	\right]	}.
\end{equation}

Calculation of vector and tensor integrals does not change the main idea in Eqs. (\ref{eq:simpl}-\ref{result})
\begin{eqnarray}
	I_\mu &=& \int \frac{d p}{\pi^2} F(p^2)
	\frac{p^\mu}{ \left[
	(p+q_1)^2+m_1^2
	\right]
	\left[
	(p+q_2)^2+m_2^2
	\right]}   = \nonumber \\
	&=& - P^\mu \int_0^1 d\alpha (\alpha b_1 + (1-\alpha)b_2) \int_0^{\infty} dt  \frac{t^2}{(1+t)^3} [F(z_0)] .
\end{eqnarray}

The tensor integral
\begin{eqnarray}
	I_{\mu\nu} &=& \int \frac{d p}{\pi^2} F(p^2)
	\frac{p^\mu p^\nu}{ \left[
	(p+q_1)^2+m_1^2
	\right]
	\left[
	(p+q_2)^2+m_2^2
	\right]}    \nonumber ]\\ 
&=&
	\frac{1}{2} \delta_{\mu \nu} \int_0^1 d \alpha  
	\int_0^{\infty}dt \frac{t}{(1+t)^3} \int_0^{\infty} F[z_0+u]du
	\nonumber \\ &&
	+ P^\mu P^\nu \int_0^1 d \alpha (\alpha b_1 + (1-\alpha)b_2)^2 \int_0^{\infty} dt \frac{t^3}{(1+t)^4}  [-F'[z_0]]. 
\end{eqnarray}

The process of integration for integrals with a number of points more than two is carried out in the same way. And after applying the Feynman parametrization all results can be presented in terms  of the integrals
\begin{equation}
    I(a_i, m, n, F) = \int_0^1\lbrace d \alpha_i \rbrace \prod_i \alpha_i^{a_i}\int_0^\infty dt\frac{t^m}{(1+t)^n} [F(z_0)].
\end{equation}

\section*{Appendix B: three point integrals}

 During the calculation of triangle diagrams set of integrals appears:
\begin{eqnarray*}\label{eq:integralI3}
  I(q_1,q_2,q_3) = \int \frac{d^4p}{\pi^2} F(p^2)
  \frac{\lbrace 1, p^\mu, p^{\mu\nu},p^{\mu\nu\rho}\rbrace}{(p_1^2+m_1^2)(p_2^2+m_2^2)(p_3^2+m_3^2)},
\end{eqnarray*}
where $p_1= p +q_1$ and $p_2= p + q_2$,  $p_3= p + q_3$ with  
$q_i$ defined from the kinematics of the process. The function $F(p^2)$ contains the sum of the vertex functions $\sum\varphi(p^2/\Lambda_i^2)$. The procedure of integration is the same as in Appendix A and the results are presented below with notation $D_i = ((p+q_i)^2 + m_i^2)$:
\begin{eqnarray}
    I_3 &=& \int \frac{d p}{\pi^2} F(p^2)\frac{1}{D_1 D_2 D_3} \nonumber\\
    & =& \int_0^1 \lbrace d\alpha_i\rbrace \delta(1-\sum\alpha_i)\int_0^\infty \frac{t^2}{(1+t)^2}[-F'[z_0]], \\
    I^\mu_3 &=& \int \frac{d p}{\pi^2} F(p^2)\frac{p^\mu}{D_1 D_2 D_3} \nonumber \\
    &=& -\int_0^1 \lbrace d\alpha_i\rbrace \delta(1-\sum\alpha_i) R^\mu\int_0^\infty \frac{t^3}{(1+t)^3}[-F'[z_0]], \nonumber\\
    I^{\mu\nu}_3 &=& \int \frac{d p}{\pi^2} F(p^2)\frac{p^\mu p^\nu}{D_1 D_2 D_3} \nonumber \\
    &=&\frac{1}{2}\delta_{\mu\nu}\int_0^1 \lbrace d\alpha_i\rbrace \delta(1-\sum\alpha_i)\int_0^\infty \frac{t^2}{(1+t)^3} F[z_0]\nonumber\\
    &+& \int_0^1 \lbrace d\alpha_i\rbrace \delta(1-\sum\alpha_i) R^\mu R^\nu\int_0^\infty \frac{t^4}{(1+t)^4}[-F'[z_0]], \nonumber\\
     I^{\mu\nu\rho}_3 &=& \int \frac{d p}{\pi^2} F(p^2)\frac{p^\mu p^\nu}{D_1 D_2 D_3}  \nonumber \\
    &=& -\frac{1}{2}\int_0^1 \lbrace d\alpha_i\rbrace \delta(1-\sum\alpha_i)(R^\mu\delta^{\nu\rho} + R^\nu\delta^{\mu\rho}+R^\rho\delta^{\mu\nu})\nonumber\\
    &&\times\int_0^\infty \frac{t^3}{(1+t)^4} F[z_0]\nonumber\\
    &-& \int_0^1 \lbrace d\alpha_i\rbrace \delta(1-\sum\alpha_i) R^\mu R^\nu R^\rho\int_0^\infty \frac{t^5}{(1+t)^5}[-F'[z_0]],\nonumber 
\end{eqnarray}
where $R = \sum \alpha_i q_i$, $D = \sum{\alpha_i (q_i^2- m_i^2)} - R^2$, $z_0$ is defined in Eq. (\ref{eq:z0}), $q_i$ depends on the kinematics of the process. For three-vertex diagram, up to three vertex functions $\displaystyle{e^{-(p^2/\Lambda_i^2)}}$ appear, and function 
\begin{equation}
    F(p^2) = \rm{exp}\left( - p^2 \sum_{n=1}^3\frac{1}{\Lambda_n}\right) = \rm{exp}\left(-\frac{p^2}{\Lambda^2}\right)
\end{equation}
with $\Lambda  =  \left(\sum_{n=1}^{3}\frac{1}{\Lambda_n}\right)^{-1/2}$ . 

%
%

\end{document}